UNIVERSIDAD POLITECNICA DE VALENCIA

PROS
Centro de Investigación en Métodos de Producción de Software

# Informe Técnico / Technical Report

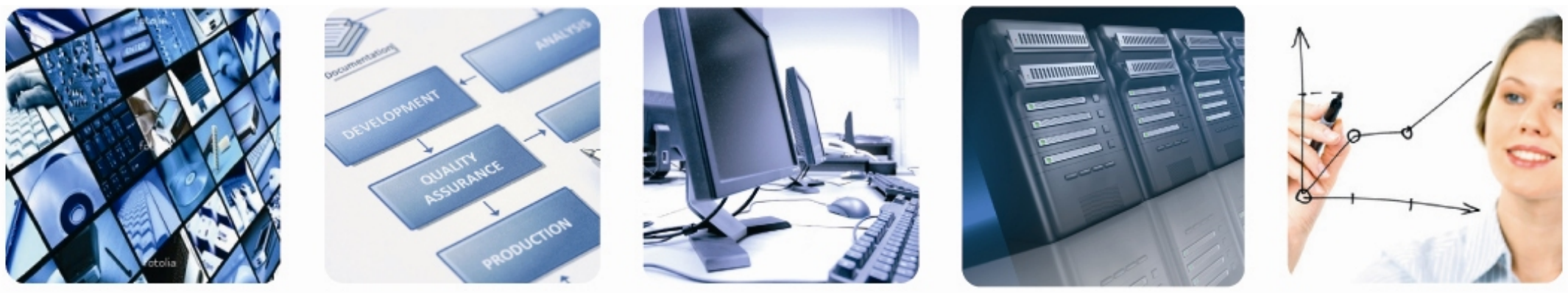

## A practical guide to Message Structures: a modelling technique for information systems analysis and design

Sergio España, Arturo González, Óscar Pastor, Marcela Ruiz

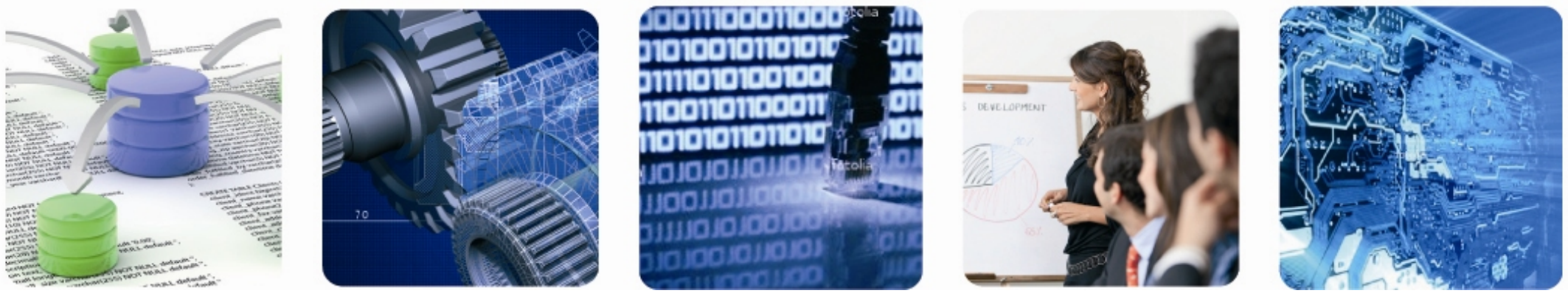

| | | | | | |
|---|---|---|---|---|---|
| Ref. #: | ProS-TR-2011-02 | | | | |
| Title: | A practical guide to Message Structures: a modelling technique for information systems analysis and design | | | | |
| Author (s): | Sergio España, Arturo González, Óscar Pastor, Marcela Ruiz | | | | |
| Corresponding author (s): | Sergio España sergio.espana@pros.upv.es<br>Marcela Ruiz lruiz@pros.upv.es | | | | |
| Document version number: | 2.0 | Final version: | No | Pages: | 30 |
| Release date: | February 2011 | | | | |
| Key words: | Information systems, analysis, design, requirements engineering, model-driven development, Message Structures, Communication Analysis | | | | |

Universidad Politécnica de Valencia · Camino de Vera s/n · Edificio 1F · 46022 Valencia Spain ·
T. +34 96 387 70 07 Ext. 83530 · M. + 34 619 543 623 · F. +34 96 387 73 59 · info@pros.upv.es · www.pros.upv.es

# MESSAGE STRUCTURES:

# A MODELLING TECHNIQUE FOR INFORMATION SYSTEMS ANALYSIS AND DESIGN

**Authors (in alphabetical order):**

**Sergio España, Arturo González, Óscar Pastor, Marcela Ruiz**

This document is born as an extended version of a paper presented in the 14th Workshop on Requirements Engineering (WER 2011). If you intend to cite Message Structures in a scientific paper, please use the following reference:

# ESTRUCTURAS DE MENSAJE:

# UNA TÉCNICA DE MODELADO PARA EL ANÁLISIS Y DISEÑO DE SISTEMAS DE INFORMACIÓN

**Autores (por orden alfabético):**

**Sergio España, Arturo González, Óscar Pastor, Marcela Ruiz**

Este documento nace como una versión extendida de un artículo presentado en el 14 Workshop en Ingeniería de Requisitos (WER 2011). En caso de querer citar las Estructuras de Mensaje en un artículo científico, se debe usar la siguiente referencia:

# TABLE OF CONTENTS

# I. ENGLISH VERSION

## 1. INTRODUCTION

Nowadays, there exists wide consensus about the importance of modelling in information systems development. However, there is still much space for improvement in model-driven software development (MDD), since many research challenges that remain open. To name a few: (i) to facilitate business process modelling from a communicational perspective [España, González et al. 2009], (ii) to provide model transformations in order to obtain the system design from the requirements models, (iii) to allow requirements traceability throughout the entire software lifecycle.

This paper presents in detail a technique for the specification of communication with the information system (IS): Message Structures[1]. Although this technique is part of Communication Analysis, a communication-oriented requirements engineering method [España, González et al. 2009], it can be used in combination with other methods and notations as well (e.g. Use Cases, Business Process Modeling Notation). Also, Message Structures can be applied solely in analysis time, solely in design time, or even throughout the entire lifecycle.

Previous works present Message Structures concisely [González, España et al. 2008; España, González et al. 2009]. This paper extends these works and makes the following contributions:

- A precise specification of the technique is provided, including definitions and examples of its concepts, its grammatical constructs, and its acquisition operations.
- Different ways if using the message structures are described, differentiating its application in analysis time (specification of business processes from a communicational perspective) and its application in design time (derivation of IS memory models and reasoning of the user interface in terms of abstract patterns).
- Two alternative technological supports for Message Structures are presented: a modelling tool that is based in Xtext and a modelling tool that is based in Eclipse Modeling Framework.

The rest of the paper is structured as follows. Section 2 presents an overview of Communication Analysis. Section 3 presents in detail the concepts related to Message Structures, and describes its grammar. Section 4 differentiates its application in analysis time and in design time. Section 5 presents the tool support. Section 6 reviews related work. Section 7 discusses some advantages and risks of using Message Structures and presents future work.

## 2. OVERVIEW OF COMMUNICATION ANALYSIS

Communication Analysis is a requirements engineering method that proposes describing business processes from a communicational perspective. The objective is to discover and describe communicative interactions between the IS and its environment. The method stems from academic research and it evolves in collaboration with industry [González 2004]. The following definitions clarify the concepts in which the proposed modelling techniques are founded.

We refer as *communicative interaction* to an interaction between actors with the purpose of exchanging information. Depending on the preeminent direction of the communication, two types of

[1] This technique has been named differently during the evolution of Communication Analysis: Data Acquisition Structures [González 2004; España 2005], Communication Structures [España, González et al. 2009]. We consider that the term Message Structures reflects their essence more appropriately.

communicative interaction are distinguished:

– *Ingoing communicative interaction*. Its main objective is to convey to the system new meaningful information; i.e. to feed the IS memory with previously unknown data that is relevant to the organisation.
– *Outgoing communicative interaction*. Its main objective is to distribute known information to users; i.e. to consult IS memory to retrieve data and present it.

The experience in development projects has shown us that ingoing communicative interactions entail more analytical complexity. Thus, we recommend analysts to focus on these interactions during analysis.

A *communicative event* is a set of actions related to information (acquisition, storage, processing, retrieval, and/or distribution), that are carried out in a complete and uninterrupted way, on the occasion of an external stimulus [González 2004]. For Communication Analysis, a communicative event is an ingoing communicative interaction that fulfils certain unity criteria (methodological guidelines that facilitate the creation of modular models) [González, España et al. 2009]. Informally, a communicative event can be seen as a business process activity.

Communication Analysis proposes specifying business processes by means of Communicative Event Diagrams, a graphical modelling technique whose notation is similar to UML Activity Diagrams. Event Specification Templates is a textual specification technique that prescribes a structure for the requirements related to a communicative event. Previous works described these techniques in detail [España, González et al. 2009]. In the following, we focus on a technique that allows the specification of the messages that are associated to business process activities.

# 3. MESSAGE STRUCTURES

Message Structures is a specification technique that allows describing, by means of structured text, the message that is associated to a communicative interaction. Although message structures can be used to specify outgoing communicative interactions, we focus on the specification of ingoing communicative interactions, given their analytical interest.

**Table 1.** Example of a message structure in analysis time

| FIELD | OP | DOMAIN | EXAMPLE VALUE |
|---|---|---|---|
| ORDER = | | | |
| < Order number + | g | number | 10352 |
| Request date + | i | date | 31-08-2009 |
| Payment type + | i | text | Cash |
| Client + | i | Client | 56746163-R, John Papiro Jr. |
| DESTINATIONS = | | | |
| { DESTINATION = | | | |
| < Address + | i | Client address | Blvd. Blue mountain, 35-14A, 2363 Toontown |
| Person in charge + | i | text | Brayden Hitchcock |
| LINES = | | | |
| { LINE = | | | |
| < Product + | i | Product | ST39455, Rounded scissors (cebra) box-100 |
| Price + | i | money | 25,40 € |
| Quantity > | i | number | 35 |
| } | | | |
| > | | | |
| } | | | |
| > | | | |

Table 1 presents an example[2] of the usage in analysis time; the ORDER message structure specifies a communicative interaction by which a client places an order[3].

[2] This particular font, the colours and the capitalisation are a non-prescriptive convention that is intended to facilitate message structure comprehension. Feel free to configure these aspects.

## 3.1. Grammatical constructs

The syntax of Message Structures can be described in terms of the following grammatical constructs.

We refer as *substructure* to an element that is part of a message structure. This way, LINE, Client and Payment type are substructures that are part of ORDER. There exist two classes of substructures: fields and complex substructures. We refer as *initial substructure* to the substructure that constitutes the first level of a message structure. For instance, ORDER=<Order number+Request date+Payment type+Client+DESTINATIONS>.

- *Field*. It is a basic informational element of the message; that which is not composed of other elements. There exist two types of fields.
  - *Data field.* It is a field that represents a piece of data with a basic domain[4]. For instance, Payment type is a data field that belongs to the message structure ORDER.
  - *Reference field.* It is a field whose domain is a type of business objects. For instance, Client references a client that is already known by the IS.
- *Complex substructure*. It is any substructure that has an internal composition. There exist three types of complex substructures.
  - *Aggregation substructure.* It specifies a composition of several substructures in a way that they remain grouped as a whole. It is represented by angle brackets < >. For instance, LINE=<Product+Price+Quantity> specifies that an order line consists of information about a product, its price and the quantity that the client requests.
  - *Iteration substructure.* It specifies a set or repetition of the substructures it contains. It is represented by curly brackets { }. For instance, an order can have several destinations and, for each destination, a set of order lines is defined. Both DESTINATIONS and LINES are iteration substructures. LINES={LINE=<Product+Price+Quantity>}
  - *Specialisation substructure.* It specifies one or more variants; that is, structural alternatives[5]. There is no example of specialisation substructure in Table 1; the message structure in Table 2 specifies that the assignment made by a student can be of type THEORY, in which case the fields Subject and Title characterise the work, or it can be of type PRACTICE, in which case the fields Programming language and Functionality characterise the work.

**Table 2.** Fragment of a message structure that includes specialisation

<table>
<tr><th>FIELD</th><th>OP</th><th>DOMAIN</th></tr>
<tr><td>< ...</td><td></td><td></td></tr>
<tr><td>Type of assignment +</td><td>i</td><td>[theo|prac]</td></tr>
<tr><td>TYPE =</td><td></td><td></td></tr>
<tr><td>[ THEORY =</td><td></td><td></td></tr>
<tr><td>< Subject +</td><td>i</td><td>Subject</td></tr>
<tr><td>Title ></td><td>i</td><td>text</td></tr>
<tr><td>| PRACTICE =</td><td></td><td></td></tr>
<tr><td>< Programming language +</td><td>i</td><td>Language</td></tr>
<tr><td>Functionality ></td><td>i</td><td>text</td></tr>
<tr><td>] + ...</td><td></td><td></td></tr>
<tr><td>></td><td></td><td></td></tr>
</table>

[3] Most of the examples in this paper are taken from a requirements model that can be found in [España, González et al. 2011].

[4] Basic domains (e.g. numbers, text) are discussed below.

[5] It is more frequent to use specialisation with two or more variants. The usage with one variant represents the optionality of that variant; that is, a message might or might not include the variant.

**Table 3.** EBNF grammar of Message Structures[6]

```
message structure
= structure name, ’=’, initial substructure;
initial substructure
= aggregation substructure | iteration substructure;
aggregation substructure
= ’<’, substructure list, ’>’;
iteration substructure
= ’{’, substructure list, ’}’;
specialisation substructure
= ’[’, substructure list,{ ’|’, substructure list },’]’;
substructure list
= substructure, { ’+’, substructure };
complex substructure
= aggregation substructure | iteration substructure
| specialisation substructure;
substructure
= substructure name, ’=’, complex substructure | field;
```

For greater disambiguation, Table 3 presents the grammatical constructs of Message Structures using the Extended Backus-Naur Form notation (EBNF) [ISO/IEC 1996].

In practice, the syntax is more flexible. The syntax of Message Structures allows creating equivalent structures by means of the following mechanisms: (i) The names of complex substructures can be omitted. (ii) An iteration substructure also aggregates its own content (there is an aggregation substructure that can be made explicit of it can be left implicit). (iii) Similarly, each variant of a specialisation substructure also has an implicit aggregation. This ‘syntactic sugar’ allows adapting the notation to project contingencies and it facilitates the usage of the technique. Table 4 illustrates these mechanisms by presenting four message structures that are semantically equivalent.

**Table 4.** Four equivalent message structures

| | | | |
|---|---|---|---|
| A = | A = | A = | A = |
| < a + | < a + | < a + | < a + |
| b + | b + | b + | b + |
| C = | { D = | C = | { e + |
| { D = | < e + | { e + | f + |
| < e + | f + | f + | g |
| f + | g | g | } |
| g | > | } | > |
| > | } | > | |
| } | > | | |
| > | | | |

## 3.2. Field specification

To characterise a field, the following properties can be specified:

- *Name*. Each field must have a significant name (e.g. Request date).
- *Acquisition operation*. It specifies the origin of the information that the field represents.
    - *Input* i. The information of the field is provided by the primary actor.
    - *Generation* g. The IS can automatically generate the information of the field.
    - *Derivation* d. The information of the field is already known by the IS and, therefore, it can be derived from its memory; that is, it was previously communicated in a preceding communicative event. This operation can have an associated derivation formula.
- *Domain*. It specifies the type of information the field contains.
- *Example.* An example of a value for the field, provided by the organisation.
- *Description.* An explanation that helps the reader to understand the field meaning.

[6] The elements `structure name`, `substructure name` and `field` are character strings.

- *Label*. A brief text that describes the field when shown in a graphical interface.
- *Link with memory*. It specifies the correspondence between the field and a database table column or a class diagram attribute.
- *Compulsoriness*. It specifies whether the message field is mandatory or not. It is possible to specify that the field is not compulsory by using a one-variant specialisation (e.g. [a]).
- *Initialisation*. The value that the field is given by default can be specified by means of a function or a derivation formula.
- *Visibility*. It specifies whether the field is visible in a graphical user interface form.

It is recommended to lay the fields out vertically and to specify the field properties horizontally (by means of columns). For reasons of space, the description of the fields can be done in a separate table. Message Structures can be extended with other field properties that a method designer or an analyst deem appropriate. However, as discussed below, not all properties are convenient at analysis time.

# 4. USAGES OF MESSAGE STRUCTURES

Message Structures can be applied for different purposes (from software development to adaptive maintenance) and in different stages of the software development life cycle (e.g. analysis, design). Depending on whether they are used in analysis or design time, syntactic and pragmatic differences have to be taken into account. Table 5 presents recommendations on the usage of field properties, depending on the development stage in which Message Structures are used.

**Table 5.** Applicability of field properties to development stage

| | | Name | Acquisition operation | | | Domain | Example | Description | Label | Link with memory | Compulsoriness | Initialisation | Visibility |
|---|---|---|---|---|---|---|---|---|---|---|---|---|---|
| | | | i | g | d | | | | | | | | |
| Analysis | | ++ | ++ | ++ | -- | ++ | ++ | ++ | -- | -- | -- | -- | -- |
| Design | Memory | ++ | ++ | ++ | ++ | ++ | ++ | ++ | - | ++ | + | - | - |
| | Interface | ++ | ++ | ++ | ++ | ++ | ++ | ++ | ++ | ++ | ++ | ++ | + |

LEGEND: ++ highly recommended + recommended - not recommended -- discouraged

## 4.1. Creation and usage of Message Structures in analysis time

In analysis time, Message Structures allow specifying in detail the communicative interactions that take place in the organisational work practice. This way, they offer a communicational perspective for business process modelling and they act as requirements for the IS. In the context of Communication Analysis, the new meaningful information that is conveyed to the IS in each communicative event is specified by means of a message structure.

In the following, we enumerate some sources of information and techniques for acquiring information and analysing the messages exchanged with the IS.

*Organisational actors* play an important role in IS analysis, since they know organisational work practice first-hand. The analyst will employ elicitation techniques such as interviews or JAD sessions [August 1991]. It is crucial to ask the proper questions so as to define which information is conveyed in each communicative event, as well as to distinguish new information from derived information.

*Business forms* are a technological support for communicative interactions and, therefore, they are a major source for analysis. In this sense, the user interface screens from pre-existing software are equivalent. Forms can be used for entering information (input forms), for presenting data (output forms), or for both purposes. In analysis time, input forms allow to identify communicative events that convey new information to the IS. The analyst has to carry on, along with the users, the following investigations:

– Whether the form is filled in one go or iteratively in different moments in time; the corresponding communicative events are identified. For instance, the client order form shown in Fig. 1 is affected by more than one communicative event[3] [España, González et al. 2011]: the request of the order, the assignment to a supplier, and the supplier response.
– Which is the temporal order in which communicative events occur.
– Which are the primary actors of the communicative events (those that are the source of the information the form is filled which).
– What message is conveyed in each communicative event (the fields of the form that are affected by the event). Observe that the message structure in Table 1 does not include fields such as Supplier or Planned delivery date, which correspond to later communicative events.

If the organisation has *previous business process specifications* or quality procedures, then this documentation can also be used as input for the analysis.

Using these sources and techniques, analysts identify communicative events and they specify, by means of message structures, the new meaningful information communicated in each event. This information is mainly provided by the primary actor; in case the information of a field is provided by the primary actor, then it is attached an input *acquisition operation* (e.g. the quantity of items of a certain product that the client requests: Quantity i). Some informational elements may not be provided by the primary actor, but generated by the IS itself; in that case the acquisition operation is generation (e.g. order numbers, as well as invoice numbers, are usually pre-printed in form booklets: Order number g).

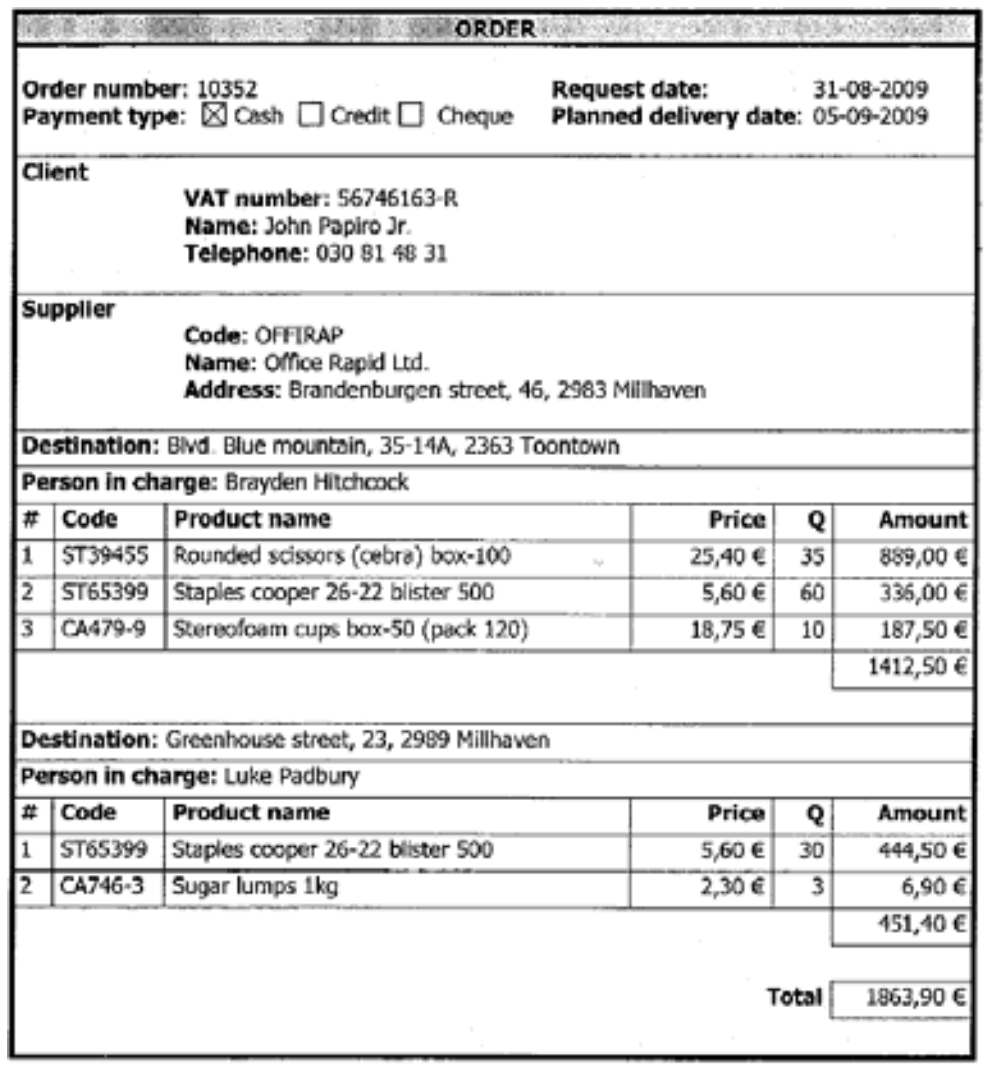

**ORDER**

**Order number:** 10352 **Request date:** 31-08-2009
**Payment type:** ☒ Cash ☐ Credit ☐ Cheque **Planned delivery date:** 05-09-2009

**Client**
**VAT number:** 56746163-R
**Name:** John Papiro Jr.
**Telephone:** 030 81 48 31

**Supplier**
**Code:** OFFIRAP
**Name:** Office Rapid Ltd.
**Address:** Brandenburgen street, 46, 2983 Millhaven

**Destination:** Blvd. Blue mountain, 35-14A, 2363 Toontown
**Person in charge:** Brayden Hitchcock

| # | Code | Product name | Price | Q | Amount |
|---|---|---|---|---|---|
| 1 | ST39455 | Rounded scissors (cebra) box-100 | 25,40 € | 35 | 889,00 € |
| 2 | ST65399 | Staples cooper 26-22 blister 500 | 5,60 € | 60 | 336,00 € |
| 3 | CA479-9 | Stereofoam cups box-50 (pack 120) | 18,75 € | 10 | 187,50 € |
| | | | | | 1412,50 € |

**Destination:** Greenhouse street, 23, 2989 Millhaven
**Person in charge:** Luke Padbury

| # | Code | Product name | Price | Q | Amount |
|---|---|---|---|---|---|
| 1 | ST65399 | Staples cooper 26-22 blister 500 | 5,60 € | 30 | 444,50 € |
| 2 | CA746-3 | Sugar lumps 1kg | 2,30 € | 3 | 6,90 € |
| | | | | | 451,40 € |

**Total** 1863,90 €

**Fig. 1.** Example of an order form

With regards to the *domain*, in analysis time it is advised to use five basic types for data fields: text, number, money, date, time. For instance, the domain of Quantity is number. For reference fields, the domain is the type of business object. For instance, the field Client identifies a client that has already been registered in the IS; thus, its domain is Client (similarly, Address refers to a client address and Product identifies a product of the catalogue). Also, if the values that the field can take are restricted to a predefined set and there is not much prospect that the set will be updated, then the domain can be expressed by means of a specialisation substructure (e.g. the domain of Type of work in the example in Table 2 is [theo|prac]).

Also, in order to enhance the comprehensibility of analysis specifications, an *example* is provided for each field (e.g. Request date, 31-08-2009), as well as a *description*.

In analysis time, it is discouraged to specify derived fields (data fields whose acquisition operation is d), as well as any other field property that is an aspect of design (see Table 5). Derived fields do not convey new information and they contribute to an unnecessary over-specification. It is convenient to postpone specifying derivation until the design stage. In order to avoid specifying derived fields, each time it is needed to identify a business object that is already known by the IS (e.g. a client), the guideline is to include a reference field (e.g. Client i Client), and to avoid including the information that characterises the object (e.g. the client name). Equally, the analyst will avoid including fields whose content can be derived from the rest of the information in the message (e.g. total amounts such as Amount or Total).

In any case, notice that the message structure in Table 1 includes a field named Price that corresponds to the price of the product. In fact, this field is infringing the above-mentioned methodological guidelines. The field actually refers to information that is already known by the IS (the price is part of the company catalogue). The analyst has included this field with a design solution in mind: the IS is intended to register the Price of the product at the time of the order request so as to guarantee that, even though the prices are yearly updated, the information on the orders and invoices will remain unchanged. However, this is only one possible design solution among others (e.g. a archival storage of prices). In case analysts decide to include design aspects in analysis time, then they should be well aware and justify their decision.

## 4.2. Usage of Message Structures in design time

In design time, Message Structures allow establishing the traceability between analysis documentation, the specification of the IS memory (e.g. by means of a class diagram or a relational database schema), and the specification of the user interface. Moreover, it is possible to define techniques for deriving the memory if the IS from requirements models, as well as techniques for systematically reasoning the interface design.

The summarised procedure for the derivation of the IS memory is as follows[7]. First the communicative events are sorted according to their temporal precedence. Then the message structure of each event is processed in order to obtain a class diagram view. This way, the complete class diagram is iteratively created by integrating the class diagram views that correspond to all the communicative events. Fig. 2 (right-hand side) shows the derivation of the class diagram view that corresponds to a communicative event in which a client places an order. A more detailed and bigger example is available online [España, González et al. 2011].

The summarised procedure to reason the user interface is as follows. First the interface style manual. Then the editing environments are identified (i.e. sets of forms or interface screens that support a set of editorially-compatible communicative events. Next, the message structures are fragmented (e.g. normalising them in first normal form) and the fragments are assigned to abstract interface structures (e.g. registry, set of registries). The abstract interface structures are encapsulated in forms. Each form is specified in detail, establishing the possible interaction and the editing facilities (filters, order criteria). The behaviour of the interface can be specified by means of trigger tables. Lastly, additional listings and printouts are specified. Fig. 2 (left-hand side) shows how the IS interface is designed in terms of abstract interface patterns. Methodological guidelines are described in detail in [España 2005].

[7] We describe the derivation of class diagrams because this derivation technique is part of ongoing research. An analogous argumentation can be made for relational schemas.

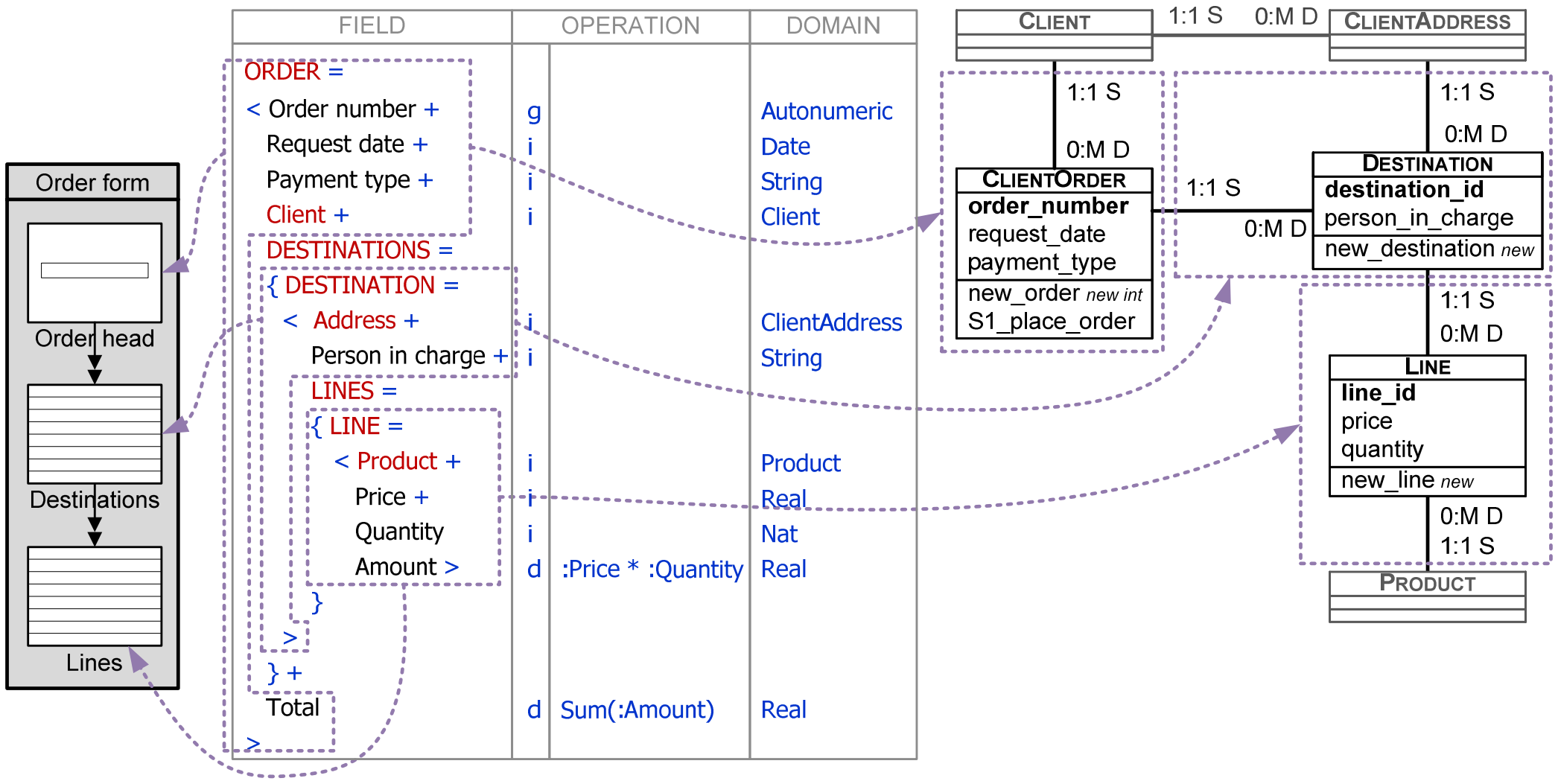

**Fig. 2.** Derivation of an interface design and a class diagram view that correspond to a communicative event in which an order is placed

In design time, it is usual to specify derived fields (e.g. the total amount of each order line Amount d (:Price * :Quantity)). Other properties that specify aspects of design are also specified in this stage (e.g. the specification of the data field Request date could also include a formula that defines its initialisation value: today()).

# 5. TECHNOLOGICAL SUPPORT FOR MESSAGE STRUCTURES

Model-driven software development (MDD) has become a reality [Pastor and Molina 2007]. However, the community lacks a collection of well integrated CASE tools that cover all the lifecycle, from requirements engineering to automatic software code generation, and that take advantage of model transformations in all the stages. This section presents the modelling tools that are being developed to support Message Structures. These supports extend a CASE tool under development that aims for the integration of Communication Analysis in MDD environments[Ruiz, España et al. 2010]. Two alternative development environments have been chosen; namely, Xtext and the Eclipse Modeling Framework.

## 5.1. Support to Message Structures with Xtext

Message Structures is a modelling language based on structured text, that can be specified using the extended Backus-Naur Form notation (see Table 3). This characteristic facilitates the development of a domain-specific language (DSL) tool. Fig. 3.a shows the Message Structure grammar as defined in the Xtext environment, an Eclipse-based environment for the development of textual DSLs [Behrens, Clay et al. 2010]. This environment allows the automatic generation of textual editors for the defined DSLs. Fig. 3.b shows the specification of the message structure ORDER, using the Xtext tool. An advantage of this environment is that it can be integrated with other Eclipse-based modelling tools.

```
grammar org.xtext.example.mydsl.CAMS with
org.eclipse.xtext.common.Terminals
generate cAMS
"http://www.xtext.org/example/mydsl/CAMS"
MessageStruc:
strucName +=StrucName
(initialSubstruc +=InitialSubstruc);
StrucName:
strucName=ID '=';
InitialSubstr:
(aggregationSubstruc +=AggregationSubstruc) |
(iterationSubstruc +=IterationSubstruc);
AggregationSubstruc:
'<'(substrucList +=SubstrucList)'>';
IterationSubstruc:
'{'(substrucList +=SubstrucList)'}';
SpecialisationSubstruc:
'['(substrucList +=SubstrucList)
('|' (substrucList +=SubstrucList))']';
SubstrucList:
(substruc+=Substruc) ('+'(substruc+=Substruc))*;
Substruc:
(field +=Field) |
substrucName=ID'='
(complexSubstruc+=ComplexSubstruc);
Field:
fieldName=ID;
ComplexSubstruc:
(aggregationSubstruc+=AggregationSubstruc)|
(iterationSubstruc+=IterationSubstruc)|
(specialisationSubstruc+=SpecialisationSubstruc);
```

**a)** DSL definition in Xtext for Message Structures

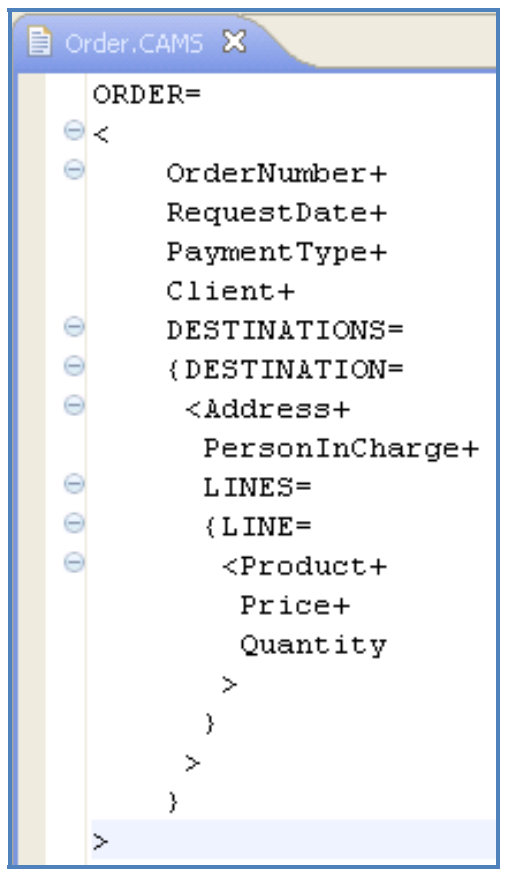

**b)** Example of a message structure

**Fig. 3.** Support to Message Structures with the Xtext environment

## 5.2. Support to Message Structures with Eclipse Modeling Framework

The MDD paradigm can add value to requirements models: the potential to derive from them conceptual models that can later be used for automatic code generation. With this aim in mind, [Ruiz, España et al. 2010] defines a process to integrate Communication Analysis in a MDD environment and it presents a metamodel that specifies part of the method. This metamodel was created using Eclipse Modeling Framework (EMF) (first a class diagram was designed using UML2 Tools and then an Ecore metamodel was generated).

This paper presents an extension of the metamodel for Communication Analysis that allows modelling Message Structures. In Fig. 4, pre-existing metaclasses have a gray border, whereas added metaclasses have a black border.

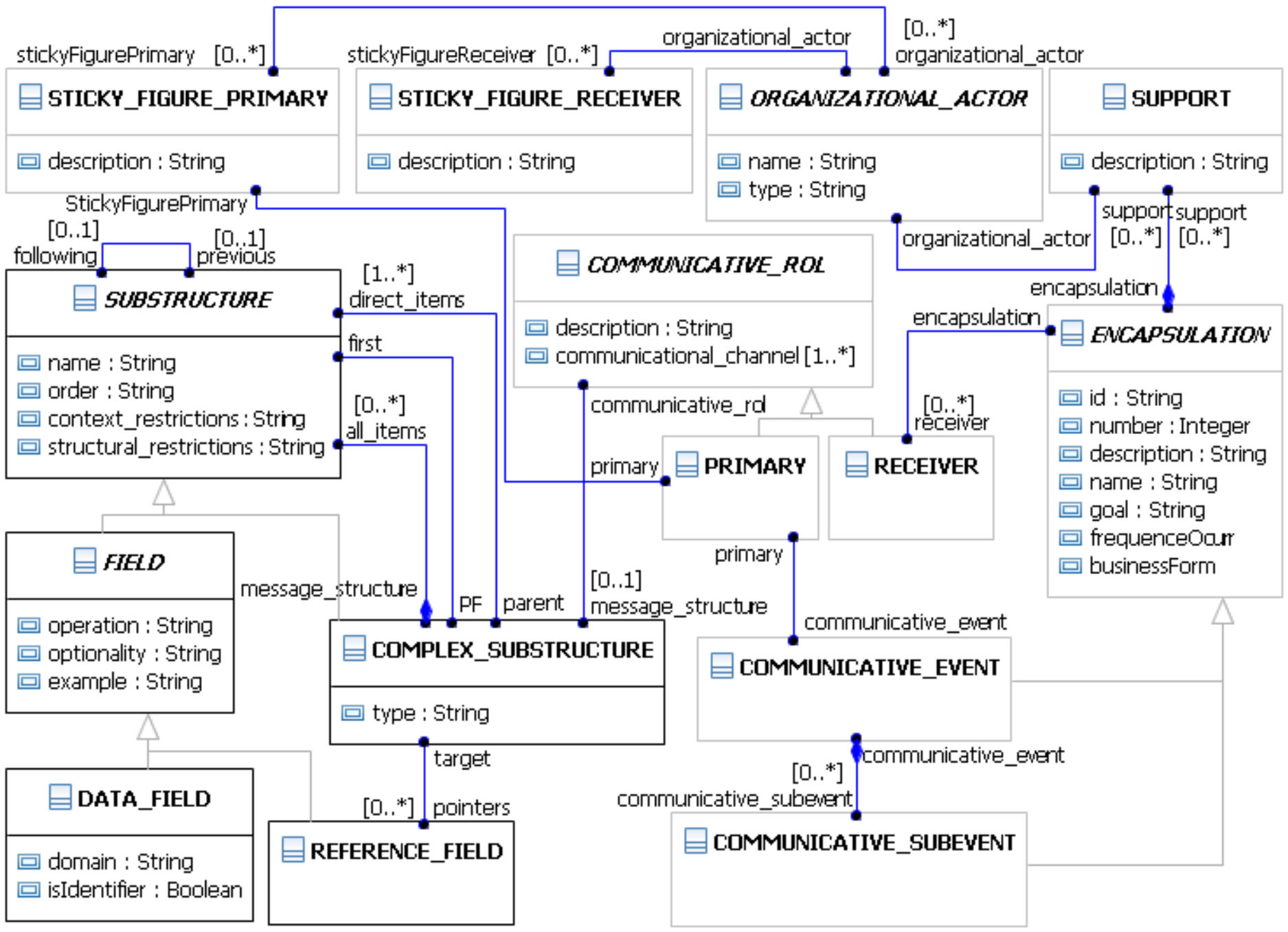

**Fig. 4.** Support to Message Structures with Eclipse Modeling Framework (partial view of the extended metamodel)

Fig. 5 shows the message structure as an instantiation of the metamodel, in the form of an Ecore tree. The tree graphically represents the composition of complex substructures, leaving the operators = and + implicit. The type of each substructure is stored in the property type of the metaclass COMPLEX_SUBSTRUCTURE (e.g. the tab named Properties shows that the complex substructure DESTINATIONS is of type iteration).

On the one hand, the implementation in Xtext ensures the compliance with the EBNF grammar for Message Structures and it offers an editorial environment that is more efficient and usable. On the other hand, the implementation in EMF extends the CASE tool for Communication Analysis; moreover, its Ecore metamodel offers the possibility of defining model to model transformations using languages such as ATL Transformation Language (ATL[Jouault and Kurtev 2005]) or Query/View/Transformation (QVT [OMG 2008]). In any case, both implementation approaches are complementary, since both environments (Xtext and EMF) can be integrated (this is planned as future work).

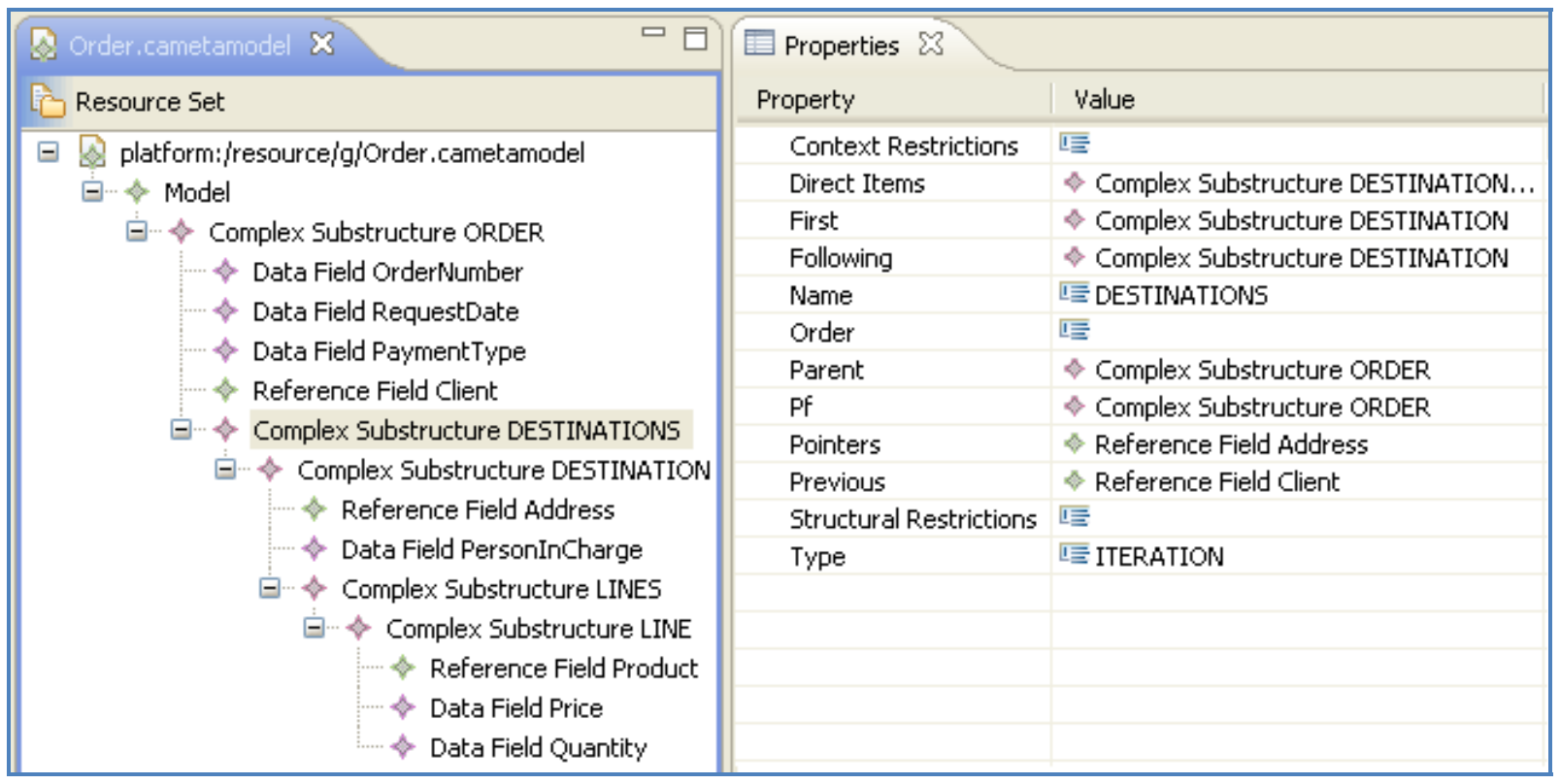

**Fig. 5.** Example of message structure supported by the EMF environment

# 6. RELATED WORKS

There exist several communicational approaches to information systems analysis; e.g. Action Workflow [Medina-Mora, Winograd et al. 1992], SAMPO [Auramäki, Lehtinen et al. 1988], Business Action Theory [Goldkuhl 1996], DEMO [Dietz 1999], Cronholm and Goldkunhl's Communication Analysis [Cronholm and Goldkuhl 2004], SANP [Chang and Woo 1994], Organisational Semiotics [Stamper 1997]. We share with them the communicational perspective and many foundations borrowed from communication theory. However, Communication Analysis differs in several conceptual foundations, in the underlying requirements structure, and in the business process modularity guidelines. For a more detailed comparison see [España, González et al. 2009]. Moreover, unlike the above-mentioned approaches, Communication Analysis places the emphasis on specifying the messages that correspond to ingoing communicative interactions; to that end, we propose Message Structures.

A notable ancestor of Message Structures is Backus-Naur Form (BNF). BNF is a notation for context-free grammars that was proposed during the development of the compiler Algol 60 [Backus, Bauer et al. 1963]. The grammatical constructs are the sequence, represented by contiguity, and the alternative, represented by a vertical bar |. Later extensions incorporate the curly brackets { } for repetitions, and the square brackets [ ] for the alternative. Structured Analysis adapts BNF for the specification of data structures [DeMarco 1979]. Fortuna et al. [Fortuna and Borges 2005] propose extending UML Use Cases with a BNF-based notation to specify informational interfaces, with a similar intention to Message Structures.

However, no previous notation includes an operator to explicitly parenthesise sequences. This prevents the analyst from including substructures within substructures and forces the fragmentation of complex substructures. For instance, the first structure (a) presents an ambiguity that can only be avoided by fragmenting the structure in two parts (b) or, as proposed by Communication Analysis, parenthesising the aggregation (c). An advantage of (c) over (b) is that (c) preserves the unity of the message.

a) Vehicle=NumberPlate+Brand+Model+Motor=CubicCapacity+Valves+Fuel+Colour
b) Vehicle=NumberPlate+Brand+Model+Motor+Colour
   Motor=CubicCapacity+Valves+Fuel
c) Vehicle=NumberPlate+Brand+Model+Motor=<CubiCapacity+Valves+Fuel>+Colour

Other techniques that allow modelling the interaction between the organisational actors and the IS can be used instead of Message Structures; for instance, UML Sequence Diagrams and ConcurTaskTrees [Panach, España et al. 2008]. However, our experience with these techniques has shown us the convenience of using a lightweight notation, such as Message Structures, which has the advantage of allowing a fast, textual specification.

# 7. DISCUSSION AND FUTURE WORK

Information systems (IS) support organisational communication. This paper presents Message Structures. Message Structures is a technique that specifies the communicative interactions among the organisational actors and the IS. Besides detailing the grammar of Message Structures, this paper also explains its uses in analysis and design time. In analysis time, message structures facilitate the identification of communicative events (business activities that provide new significant information to the IS) and complement the business process description. In design time, message structures allow deriving the IS memory and determining the interface design. In addition, this paper provides methodological guidelines and illustrative examples.

Message Structures are a part of Communication Analysis, which is a communication-oriented requirements engineering method. However, in the case that an enterprise desires to continue using a particular method, it is possible to extend it with Message Structures (this has been done for Use Cases). Since the technique is based on other well-known techniques, the adoption of Message Structures is facilitated. Note that, if Message Structures are used in combination with other techniques for the specification of the interaction (e.g. UML Sequence Diagram), redundancies may appear. In such cases, it is advised to define rules to derive some models from others or to establish procedures to verify the consistence.

Clear distinction between analysis and design is important in the use of message structures. This paper provides guidelines for the use the technique in the analysis and design phases. The use of Message Structures helps an analyst to not specify design aspects in analysis time without being conscious of it (something that can occur due to lack of experience with the technique). The risk of work overload and over-specification of the requirements model decreases with the practice, as the nuances of the technique are internalised.

Due to the fact that Message Structures is a textual notation, it offers the advantage of allowing their specification by means of a word processor or by configuring textual fields in CASE tools. However, our experiences in industrial developments, in teaching master courses in software engineering, and in laboratory experiments (e.g. [España, Condori-Fernández et al. 2010]), have convinced us to offer more versatile support. This paper presents two tools that support the technique. One tool is based on the Xtext technology. Another one is based on the Eclipse Modeling Framework. As future work, we plan to integrate both tools. This integration will take advantage of the usability of the first technology and the capacity to support model transformation of the second technology. We are currently working on deriving conceptual models from requirements models that include message structures. This derivation is implemented using ATL Transformation Language rules.

# II. VERSIÓN EN ESPAÑOL

## 1. INTRODUCCIÓN

Hoy en día, existe amplio consenso acerca de la importancia del modelado en el desarrollo de sistemas de información (SI). Sin embargo, todavía hay aspectos por mejorar en el área del desarrollo de software dirigido por modelos (DSDM): (i) facilitar el modelado de procesos de negocio desde una perspectiva comunicacional [España, González et al. 2009], (ii) guiar la creación de unos modelos a partir de los anteriores (p.e. de razonar el diseño en función de los requisitos), (iii) permitir la trazabilidad de requisitos a lo largo del ciclo de vida.

En este artículo presentamos con detalle una técnica para especificar la comunicación con el SI: las Estructuras de Mensaje[8]. Esta técnica nace vinculada al Análisis de Comunicaciones, un método de ingeniería de requisitos con orientación comunicativa [España, González et al. 2009], si bien se puede usar en combinación con otras notaciones (v.g. con Casos de Uso o con Business Process Modeling Notation). Además, las Estructuras de Mensaje se pueden aplicar aisladamente en tiempo de análisis, de diseño, o a lo largo de todo el ciclo de vida.

Trabajos anteriores introducen de forma concisa las Estructuras de Mensaje [González, España et al. 2008; España, González et al. 2009]. Este artículo extiende estos trabajos y realiza las siguientes contribuciones:

- Se ofrece una especificación precisa de la técnica, incluyendo definiciones y ejemplos de conceptos, constructores gramaticales y operaciones de adquisición.
- Se describen los usos de las estructuras de mensaje, diferenciando su uso en tiempo de análisis (especificación de procesos de negocio desde un punto de vista comunicacional) de sus usos en tiempo de diseño (derivación de los modelos de memoria del sistema de información y razonamiento de la interfaz de usuario).
- Se presentan dos ejemplos alternativos de soporte tecnológico para la especificación de Estructuras de Mensaje: un modelador basado en Xtext y un modelador basado en Eclipse Modeling Framework.

La estructura del artículo es la siguiente. La Sección 2 presenta resumidamente el Análisis de Comunicaciones. La Sección 3 presenta con detalle los conceptos relacionados con las Estructuras de Mensaje y describe su gramática. La Sección 4 diferencia los usos en tiempo de análisis y tiempo de diseño. La Sección 5 presenta las herramientas de soporte. La Sección 6 revisa trabajos relacionados. La Sección 7 discute las ventajas y los riesgos de las Estructuras de Mensaje, y enumera los trabajos futuros.

## 2. PERSPECTIVA GENERAL DEL ANÁLISIS DE COMUNICACIONES

El Análisis de Comunicaciones es un método de ingeniería de requisitos que propone describir los procesos de negocio desde una perspectiva comunicativa. El objetivo es descubrir y describir las interacciones comunicativas entre el SI y su entorno. Nace de la investigación académica y evoluciona en colaboración con la industria [González 2004]. Las siguientes definiciones clarifican los conceptos que fundamentan las técnicas de modelado propuestas.

---

[8] Esta técnica ha recibido otros nombres durante la evolución del Análisis de Comunicaciones: Estructuras de Adquisición de Datos [González 2004; España 2005], Estructuras de Comunicación [España, González et al. 2009]. Consideramos que el término Estructuras de Mensaje refleja de manera más adecuada su naturaleza.

Se define *interacción comunicativa* como la interacción entre actores con el objetivo de intercambiar información. Dependiendo de la dirección preponderante de la comunicación, se distinguen dos tipos de interacción comunicativa:

– *Interacción comunicativa entrante*. Su objetivo fundamental es aportar al sistema nueva información significativa; es decir, alimenta la memoria del SI con datos relevantes para la organización, previamente desconocidos.
– *Interacción comunicativa saliente*. Su objetivo fundamental es distribuir datos conocidos a los usuarios; es decir, consulta la memoria del sistema para recuperar datos y presentarlos a los usuarios.

La experiencia en proyectos de desarrollo nos ha enseñado que las interacciones comunicativas entrantes comportan mayor complejidad analítica. Por lo tanto, se aconseja al analista centrar la atención en éstas durante el análisis.

Un *suceso comunicativo* es un conjunto de acciones relacionadas con la información (adquisición, almacenamiento, proceso, recuperación y/o distribución), que se llevan a cabo de manera completa y sin interrupciones, con motivo de un estímulo externo [González 2004]. Para el Análisis de Comunicaciones, un suceso comunicativo es una interacción comunicativa entrante que cumple ciertos criterios de unidad (guías metodológicas que facilitan la creación de modelos modulares) [González, España et al. 2009].

El Análisis de Comunicaciones propone especificar los procesos de negocio mediante *Diagramas de Sucesos Comunicativos*, una técnica de modelado gráfico de notación semejante a los Diagramas de Actividad. Las *Plantillas de Especificación de Sucesos* son una técnica de especificación textual que prescribe una estructura para los requisitos asociados a un suceso comunicativo. Estas técnicas se describen con más detalle en trabajos previos [España, González et al. 2009]. En lo sucesivo nos centramos en una técnica que permite la especificación del mensaje asociado a cada interacción comunicativa.

# 3. ESTRUCTURAS DE MENSAJE

Estructuras de Mensaje es una técnica de especificación que permite describir, mediante texto estructurado, el mensaje asociado a una interacción comunicativa. Si bien las estructuras de mensaje pueden usarse para especificar interacciones comunicativas salientes, nos centramos principalmente en la especificación de las interacciones comunicativas entrantes, por el interés analítico que conllevan.

La Table 6 presenta un ejemplo[9] de la estructura de mensaje ORDER, que especifica una interacción comunicativa entrante por la cual un cliente realiza un pedido[10].

[9] La tipografía, los colores y el uso de mayúsculas son una convención no prescriptiva que está destinada a facilitar la comprensión de las estructuras de mensaje.

[10] La mayor parte de los ejemplos de este artículo pertenecen a un modelo de requisitos que puede encontrarse en [España, González et al. 2011].

**Table 6.** Ejemplo de una estructura de mensaje en tiempo de análisis

| FIELD | OP | DOMAIN | EXAMPLE VALUE |
|---|---|---|---|
| ORDER = | | | |
| < Order number + | g | number | 10352 |
| Request date + | i | date | 31-08-2009 |
| Payment type + | i | text | Cash |
| Client + | i | Client | 56746163-R, John Papiro Jr. |
| DESTINATIONS = | | | |
| { DESTINATION = | | | |
| < Address + | i | Client address | Blvd. Blue mountain, 35-14A, 2363 Toontown |
| Person in charge + | i | text | Brayden Hitchcock |
| LINES = | | | |
| { LINE = | | | |
| < Product + | i | Product | ST39455, Rounded scissors (cebra) box-100 |
| Price + | i | money | 25,40 € |
| Quantity > | i | number | 35 |
| } | | | |
| > | | | |
| } | | | |
| > | | | |

## 3.1. Constructores gramaticales

La sintaxis de las estructuras de mensaje puede ser descrita en términos de los siguientes constructores gramaticales.

Llamamos *subestructura* a un elemento constitutivo de una estructura de mensaje. De esta manera, LINE, Client y Payment type son subestructuras de ORDER. Existen dos tipos de subestructuras: campos y subestructuras complejas. Llamamos *subestructura inicial* a la subestructura que constituye el primer nivel de una estructura de mensaje. Por ejemplo, ORDER=<Order number+Request date+Payment type+Client+DESTINATIONS>.

- *Campo*. Se trata de un elemento informacional básico del mensaje; aquel que no está compuesto de otros elementos. Existen dos tipos de campos.
  - *Campo de datos.* Especifica un campo que representa un dato cuyo dominio es básico[11]. Por ejemplo, Payment type es un campo de datos textual, de la estructura de mensaje ORDER.
  - *Campo de referencia.* Especifica un campo cuyo dominio es un objeto de negocio. Por ejemplo, Client referencia a un cliente ya conocido por el SI.
- *Subestructura compleja*. Se trata de cualquier subestructura que tiene una composición interna. Existen tres tipos de subestructura compleja.
  - *Subestructura de agregación.* Especifica una composición de varias subestructuras de manera que quedan agrupadas. Se representa mediante paréntesis angulares < >. Por ejemplo, LINE=<Product+Price+Quantity> especifica que una línea de pedido consiste de información sobre un producto, su precio, y la cantidad solicitada por el cliente.
  - *Subestructura de iteración.* Especifica un conjunto o repetición de aquellas subestructuras que contiene. Se representa mediante llaves { }. Por ejemplo, un pedido puede tener varios destinos y, para cada destino, se define un conjunto de líneas de pedido. Tanto DESTINATIONS como LINES son subestructuras de iteración. LINES={LINE=<Product+Price+Quantity>}
  - *Subestructura de especialización.* Especifica una o más variantes; es decir, alternativas estructurales[12]. No hay ningún ejemplo de subestructura de especialización en la Table 6; la estructura de mensaje de la Table 7 especifica que el trabajo de un estudiante puede ser bien de tipo THEORY, en cuyo caso los campos Subject y Title caracterizan el trabajo, o bien de tipo PRACTICE, en cuyo caso lo hacen los campos Programming language y Functionality.

[11] Los dominios básicos (v.g. números, texto) se describen con mayor detalle más adelante.

**Table 7.** Fragmento de una estructura de mensaje que incluye especialización

<table>
<tr><th>FIELD</th><th>OP</th><th>DOMAIN</th></tr>
<tr><td>< …</td><td></td><td></td></tr>
<tr><td>Type of work +</td><td>i</td><td>[theo|prac]</td></tr>
<tr><td>TYPE =</td><td></td><td></td></tr>
<tr><td>[ THEORY =</td><td></td><td></td></tr>
<tr><td>< Subject +</td><td>i</td><td>Subject</td></tr>
<tr><td>Title ></td><td>i</td><td>text</td></tr>
<tr><td>| PRACTICE =</td><td></td><td></td></tr>
<tr><td>< Programming language</td><td>i</td><td>Language</td></tr>
<tr><td>+</td><td>i</td><td>text</td></tr>
<tr><td>Functionality ></td><td></td><td></td></tr>
<tr><td>] + …</td><td></td><td></td></tr>
<tr><td>></td><td></td><td></td></tr>
</table>

Para mayor desambiguación, la Table 8 presenta los constructores gramaticales de las EM usando la notación Backus-Naur Form extendida (EBNF) [ISO/IEC 1996].

**Table 8.** Grámatica EBNF de las Estructuras de Mensaje[13]

```
message structure
= structure name, ’=’, initial substructure;
initial substructure
= aggregation substructure | iteration substructure;
aggregation substructure
= ’<’, substructure list, ’>’;
iteration substructure
= ’{’, substructure list, ’}’;
specialisation substructure
= ’[’, substructure list,{ ’|’, substructure list },’]’;
substructure list
= substructure, { ’+’, substructure };
complex substructure
= aggregation substructure | iteration substructure
| specialisation substructure;
substructure
= substructure name, ’=’, complex substructure | field;
```

En la práctica, la sintaxis es más flexible. La sintaxis de las estructuras de mensaje permite construir estructuras equivalentes por medio de los siguientes mecanismos: (i) Los nombres de las subestructuras complejas pueden omitirse. (ii) Una subestructura de iteración a su vez agrega su contenido (hay una estructura de agregación que puede hacerse explícita o dejarse implícita). Las siguientes estructuras son equivalentes. (iii) Del mismo modo, cada una de las variantes alternativas de una subestructura de especialización lleva implícita una agregación. Esto permite adaptar la notación a las contingencias de cada proyecto y facilita su uso. De este modo, la Table 9 presenta cuatro estructuras de mensaje semánticamente equivalentes.

[12] Es más común el uso con dos o más variantes. El uso con una sola variante representa la opcionalidad de dicha variante; es decir, que un mensaje puede incluir o no la subestructura que contiene la variante.

[13] Los elementos `structure name`, `substructure name` y `field` son cadenas de caracteres.

**Table 9.** Cuatro estructuras de mensaje equivalentes

| | | | |
|---|---|---|---|
| A =<br>< a +<br>b +<br>C =<br>{ D =<br>< e +<br>f +<br>g<br>><br>}<br>> | A =<br>< a +<br>b +<br>{ D =<br>< e +<br>f +<br>g<br>><br>}<br>> | A =<br>< a +<br>b +<br>C =<br>{ e +<br>f +<br>g<br>}<br>> | A =<br>< a +<br>b +<br>{ e +<br>f +<br>g<br>}<br>> |

## 3.2. Especificación de campos

Para caracterizar un campo, se pueden especificar las siguientes propiedades.

- *Nombre*. Cada campo debe tener un nombre significativo (v.g. Request date).
- *Operación de adquisición*. Especifica la procedencia de la información que representa el campo.
  - *Introducción* i. La información del campo la provee el actor primario.
  - *Generación* g. La información del campo puede ser automáticamente generada por el SI.
  - *Derivación* d. La información del campo puede ser derivada de la memoria del SI porque ya se conoce; es decir, fue comunicada en un suceso comunicativo anterior. La derivación lleva asociada una fórmula de derivación.
- *Dominio*. Especifica el tipo de información que contiene el campo.
- *Ejemplo.* Un ejemplo de valor para el campo, aportado por la organización.
- *Descripción.* Una explicación que facilite comprender el significado del campo.
- *Etiqueta*. Un texto breve que describe el campo cuando se presenta en un formulario de interfaz gráfica de usuario.
- *Vinculación con memoria*. Describe la correspondencia del campo con una columna de tabla en una base de datos o con un atributo en un diagrama de clases.
- *Obligatoriedad*. Indica si el campo debe necesariamente tomar valor o no. Es posible especificar esto usando una especialización con una sola variante (v.g. [a]).
- *Inicialización*. El valor por defecto asociado a un campo se puede especificar mediante una función o una fórmula de derivación.
- *Visibilidad*. Indica si el campo es visible en un formulario de interfaz.

Se recomienda desplegar la estructura de mensaje verticalmente y especificar las propiedades de los campos horizontalmente (mediante columnas). Por motivos de espacio, se puede separar la descripción de los campos en una tabla aparte. Las Estructuras de Mensaje pueden extenderse con aquellas propiedades que un ingeniero de métodos o un analista considere oportunos. Como se discute a continuación no todas las propiedades son recomendadas en tiempo de análisis.

# 4. USOS DE LAS ESTRUCTURAS DE MENSAJE

Las Estructuras de Mensaje pueden ser empleadas en diferentes etapas del ciclo de vida de desarrollo de software. Dependiendo de si se emplean en tiempo de análisis o de diseño, existen diferencias sintácticas y pragmáticas a tener en cuenta. La Table 10 presenta recomendaciones acerca del uso de las propiedades de campos en función de la etapa del desarrollo en que se usa una estructura de mensaje.

**Table 10.** Aplicabilidad de las propiedades de los campos a la etapa de desarrollo

| | | Nombre | Operación de adquisición | | | Dominio | Ejemplo | Descripción | Etiqueta | Vinculación con memoria | Obligatoriedad | Inicialización | Visibilidad |
|---|---|---|---|---|---|---|---|---|---|---|---|---|---|
| | | | i | g | d | | | | | | | | |
| Análisis | | ++ | ++ | ++ | -- | ++ | ++ | ++ | -- | -- | -- | -- | -- |
| Diseño | Memoria | ++ | ++ | ++ | ++ | ++ | ++ | ++ | - | ++ | + | - | - |
| | Interfaz | ++ | ++ | ++ | ++ | ++ | ++ | ++ | ++ | ++ | ++ | ++ | + |

LEYENDA: ++ muy recomendado + recomendado - no recomendado -- desaconsejado

## 4.1. Creación y uso de Estructuras de Mensaje en tiempo de análisis

En tiempo de análisis, las Estructuras de Mensaje permiten especificar en detalle las interacciones comunicativas que tienen lugar en la práctica organizacional. De esta manera, ofrecen una perspectiva comunicativa al modelado de procesos de negocio y actúan como requisitos para el SI. En el contexto del Análisis de Comunicaciones, para cada suceso comunicativo debe especificarse la estructura de mensaje que describe la nueva información significativa que se aporta el SI.

A continuación se enumeran algunas fuentes de información y técnicas para obtener información y analizar los mensajes intercambiados con el SI.

Los *actores de la organización* desempeñan un papel primordial en el análisis de SI, puesto que conocen de primera mano la práctica organizacional. El analista empleará técnicas como entrevistas o sesiones JAD [August 1991]. Se realizarán las preguntas adecuadas para definir qué información se intercambia en cada suceso comunicativo y el analista distinguirá la nueva información de la información derivada.

Los *formularios del negocio* son un soporte tecnológico para las interacciones comunicativas y, por tanto, una fuente importante para el análisis. En este sentido, las pantallas del software preexistente resultan equivalentes (omitimos la argumentación para ellas). Los formularios pueden ser usados para la introducción de datos (formularios de entrada), para la presentación de datos (formularios de salida), o para ambos propósitos. En tiempo de análisis, los formularios de entrada permiten identificar sucesos comunicativos que aportan nueva información al SI. El analista debe realizar (con ayuda de los usuarios) las siguientes investigaciones:

- Si el formulario se rellena todo de una vez o de manera iterativa en diferentes momentos; se identifican los sucesos comunicativos correspondientes. Por ejemplo, el formulario de pedido que muestra la Fig. 6 es afectado por más de un suceso comunicativo[10] [España, González et al. 2011]: la solicitud de un pedido, la asignación de un proveedor y la respuesta del proveedor.
- Cuál es el orden temporal en que ocurren los sucesos comunicativos.
- Cuáles son los actores primarios de los sucesos (aquellos que son la fuente de información con la cual se rellena el formulario).
- Qué mensaje se transmite en cada suceso comunicativo (los campos del formulario que son afectados por el suceso). Obsérvese que la estructura de mensaje en la Table 6 no incluye campos como Supplier o Planned delivery date, que corresponden a sucesos comunicativos posteriores

Si existen *especificaciones previas de procesos de negocio* o de procedimientos de calidad, esta documentación también puede usarse como entrada para el análisis.

Usando estas fuentes y técnicas, el analista identificará sucesos comunicativos y especificará, mediante estructuras de mensaje, la nueva información significativa comunicada en cada suceso. Esta información la provee principalmente el actor primario; cuando así ocurre, se especifica campo con una *operación de adquisición* de introducción (v.g. la cantidad de artículos de un producto determinado que el cliente desea: Quantity i). Excepcionalmente, algunos elementos de información no son provistos por el actor primario sino que los genera el propio SI; en este caso se especifica una operación de generación (v.g. los números de pedido, así como los de factura, suelen venir pre-impresos en las libretas de formularios: Order number g).

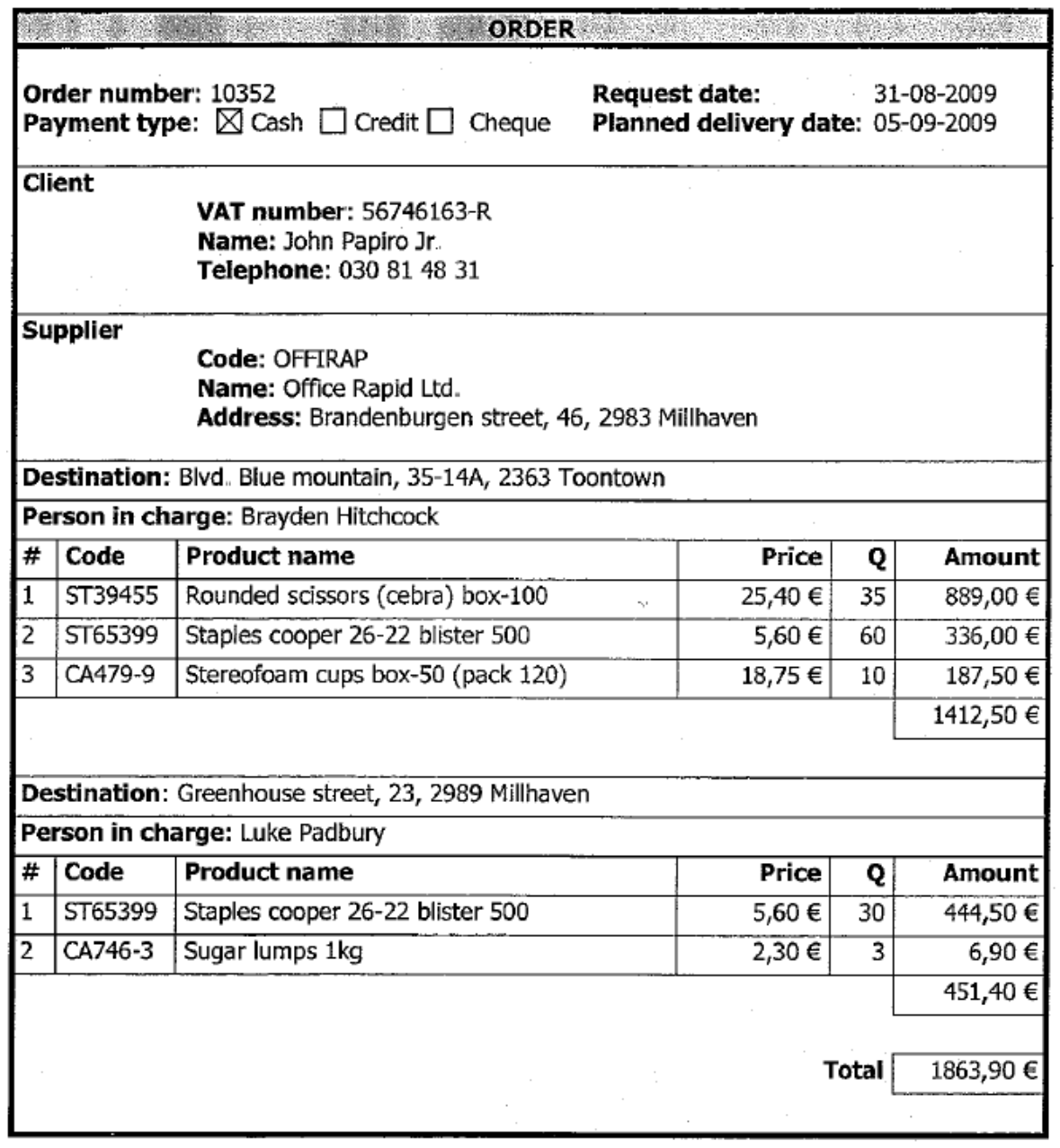
ORDER

**Order number:** 10352 **Request date:** 31-08-2009
**Payment type:** ☒ Cash ☐ Credit ☐ Cheque **Planned delivery date:** 05-09-2009

**Client**
**VAT number:** 56746163-R
**Name:** John Papiro Jr.
**Telephone:** 030 81 48 31

**Supplier**
**Code:** OFFIRAP
**Name:** Office Rapid Ltd.
**Address:** Brandenburgen street, 46, 2983 Millhaven

**Destination:** Blvd. Blue mountain, 35-14A, 2363 Toontown
**Person in charge:** Brayden Hitchcock

| # | Code | Product name | Price | Q | Amount |
|---|---|---|---|---|---|
| 1 | ST39455 | Rounded scissors (cebra) box-100 | 25,40 € | 35 | 889,00 € |
| 2 | ST65399 | Staples cooper 26-22 blister 500 | 5,60 € | 60 | 336,00 € |
| 3 | CA479-9 | Stereofoam cups box-50 (pack 120) | 18,75 € | 10 | 187,50 € |
| | | | | | 1412,50 € |

**Destination:** Greenhouse street, 23, 2989 Millhaven
**Person in charge:** Luke Padbury

| # | Code | Product name | Price | Q | Amount |
|---|---|---|---|---|---|
| 1 | ST65399 | Staples cooper 26-22 blister 500 | 5,60 € | 30 | 444,50 € |
| 2 | CA746-3 | Sugar lumps 1kg | 2,30 € | 3 | 6,90 € |
| | | | | | 451,40 € |
| | | | | **Total** | 1863,90 € |

**Fig. 6.** Ejemplo de un formulario de pedido

Respecto al *dominio*, en tiempo de análisis se recomienda el uso de cinco tipos básicos para los campos de datos: text, number, money, date, time. Por ejemplo, el dominio de Quantity es number. Para los campos de referencia, el dominio es el tipo de objeto de negocio. Por ejemplo, el campo Client identifica un cliente que ya ha sido registrado en el sistema de información; por lo tanto, su dominio es Client (de manera similar, Address se refiere a una dirección del cliente y Product identifica un producto del catálogo). Además, si los valores que puede tomar el campo están restringidos a un conjunto predefinido y sin expectativas de ser modificado, el dominio se puede expresar mediante una subestructura de especialización (v.g. el dominio de Type of work en el ejemplo de la Table 7 es [theo|prac]).

Además, para facilitar la comprensión de la especificación de análisis, se proveerá un *ejemplo* (v.g. Request date, 31-08-2009) y una *descripción* para cada campo.

En tiempo de análisis se desaconseja la especificación de campos derivados y de aquellas propiedades que constituyen aspectos de diseño (ver Table 10). Los campos derivados, no constituyen nueva información y contribuyen a una sobre-especificación innecesaria. Es conveniente relegar la especificación de la derivación a la etapa de diseño. Para evitar especificar campos derivados, cada vez que se debe identificar un objeto de negocio previamente conocido por el SI (v.g. un cliente) bastará con incluir un campo de referencia (v.g. Client i Client), y se evitará incluir la información que lo caracteriza (v.g. su nombre). Igualmente, se evitará incluir campos cuyo contenido se puede derivar del resto de la información del mensaje (v.g. importes totales como Amount o Total).

En cualquier caso, si se atiende a la estructura de mensaje de la Table 6, se puede observar que aparece el campo Price correspondiente al precio del producto. En realidad, este campo está infringiendo las guías metodológicas recién enunciadas. Se trata de información ya conocida por el SI (es parte del catálogo de la compañía). El analista ha incluido este campo para registrar el precio del producto en el momento en que se realiza el pedido y garantizar que, pese a que se actualicen los precios del catálogo anualmente, la información de los pedidos y las facturas permanecerá invariable. Sin embargo, esta solo es una solución de diseño entre varias posibles (v.g. un histórico de precios). En caso de incluir estos aspectos de diseño en tiempo de análisis, el analista debe ser plenamente consciente y justificar la decisión.

## 4.2. Uso de Estructuras de Mensaje en tiempo de diseño

En tiempo de diseño, las Estructuras de Comunicación permiten establecer la trazabilidad entre la documentación de análisis, la especificación de la memoria del sistema (por ejemplo un diagrama de clases o un esquema de bases de datos relacional) y la especificación de la interfaz de usuario. Es más, es posible definir procedimientos para derivar la memoria del sistema a partir de la especificación de análisis y para razonar sistemáticamente el diseño de la interfaz.

El procedimiento resumido para derivar la memoria del sistema es el siguiente[14]. Primero se ordenan los sucesos comunicativos en función de sus precedencias temporales. Después se trata cada suceso, procesando su estructura de mensaje correspondiente y obteniendo de ella una vista del diagrama de clases. De esta forma, el diagrama de clases completo se construye integrando incrementalmente las vistas correspondientes a todos los sucesos comunicativos. La Fig. 7 (hacia la derecha) presenta la derivación de la vista correspondiente al suceso de solicitud de un pedido. Un ejemplo más extenso y detallado está disponible online [España, González et al. 2011].

El procedimiento resumido para razonar la interfaz de usuario es el siguiente. Primero se especifican las normas de estilo de la interfaz. A continuación se identifican contextos editoriales (conjuntos de formularios o pantallas que soportarán un conjunto de sucesos comunicativos editorialmente compatibles). Después se fragmentan las estructuras de comunicación (v.g. normalizándolas en 1ª forma normal) y se asignan los fragmentos a estructuras abstractas de interfaz (v.g. registro, conjunto de registro). Se encapsulan las estructuras abstractas de interfaz en formularios. Cada formulario se especifica detalladamente, estableciendo la interacción posible y las facilidades editoriales (filtros, ordenación). El comportamiento de la interfaz se puede especificar mediante tablas de disparos. Por último, se especifican listados y formularios de salida adicionales. La Fig. 7 (hacia la izquierda) presenta el razonamiento de la interfaz del SI en términos de patrones abstractos de interfaz. Las guías metodológicas están descritas con detalle en [España 2005].

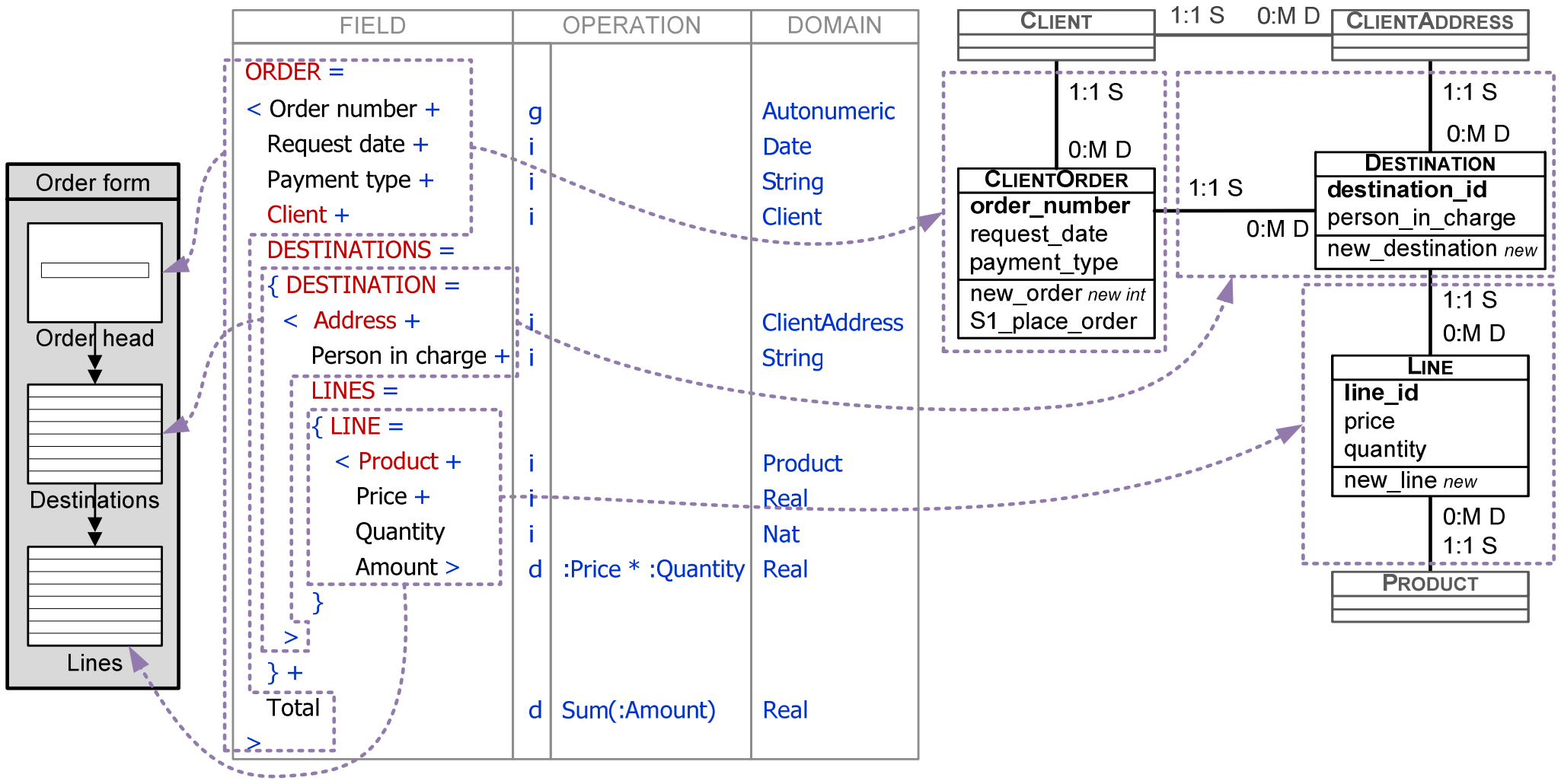

**Fig. 7.** Derivación de un diseño de interfaz y una vista de diagrama de clases correspondiente a un suceso de solicitud de un pedido

[14] Se describe la derivación de diagramas de clases porque este procedimiento es parte de los trabajos en marcha. Se puede razonar de manera análoga para esquemas relacionales.

En tiempo de diseño, es frecuente especificar campos derivados (v.g. el importe de cada línea, Amount d (:Price * :Quantity)).También se especifican propiedades de campos que especifican aspectos de diseño (v.g. el campo Request date podría tener asociada la fórmula today() para indicar que su valor de inicialización es la fecha actual).

# 5. SOPORTE TECNOLÓGICO A LAS ESTRUCTURAS DE MENSAJE

El desarrollo de software dirigido por modelos (DSDM) es ya una realidad [Pastor and Molina 2007]. Sin embargo, la comunidad adolece de un conjunto bien integrado de herramientas CASE que cubra desde los requisitos a la generación automática de código y aproveche las ventajas de las transformaciones de modelos en todas sus etapas. En esta sección se presentan los soportes tecnológicos que se están desarrollando para las Estructuras de Mensaje. Estos soportes extienden una herramienta CASE en desarrollo que persigue la integración del Análisis de Comunicaciones en entornos DSDM [Ruiz, España et al. 2010]. Se han elegido dos entornos de desarrollo alternativos: Xtext y Eclipse Modeling Framework.

## 5.1. Soporte a las Estructuras de Mensaje mediante Xtext

Las Estructuras de Mensaje son un lenguaje de texto estructurado que puede especificarse usando la notación Backus-Naur Form extendida (ver Table 8). Hemos aprovechado esta característica que facilita su desarrollo en entornos de definición de lenguajes específicos de dominio (DSL). La Fig. 8.a muestra la gramática definida en el entorno Xtext, un entorno basado en Eclipse para el desarrollo de DSL textuales [Behrens, Clay et al. 2010]. Este entorno permite la generación automática de editores textuales para los DSL definidos. La Fig. 8.b muestra la especificación de la estructura de mensaje ORDER realizada en base a la definición en Xtext. Una ventaja de este entorno es que se puede integrar con otros desarrollos basados en Eclipse.

```
grammar org.xtext.example.mydsl.CAMS with
org.eclipse.xtext.common.Terminals
generate cAMS
"http://www.xtext.org/example/mydsl/CAMS"
MessageStruc:
strucName +=StrucName
(initialSubstruc +=InitialSubstruc);
StrucName:
strucName=ID '=';
InitialSubstr:
(aggregationSubstruc +=AggregationSubstruc) |
(iterationSubstruc +=IterationSubstruc);
AggregationSubstruc:
'<'(substrucList +=SubstrucList)'>';
IterationSubstruc:
'{'(substrucList +=SubstrucList)'}';
SpecialisationSubstruc:
'['(substrucList +=SubstrucList)
('|' (substrucList +=SubstrucList))']';
SubstrucList:
(substruc+=Substruc) ('+'(substruc+=Substruc))*;
Substruc:
(field +=Field) |
substrucName=ID'='
(complexSubstruc+=ComplexSubstruc);
Field:
fieldName=ID;
ComplexSubstruc:
(aggregationSubstruc+=AggregationSubstruc)|
(iterationSubstruc+=IterationSubstruc)|
(specialisationSubstruc+=SpecialisationSubstruc);
```

**a)** DSL en Xtex para la especificación de Estructuras de Mensaje

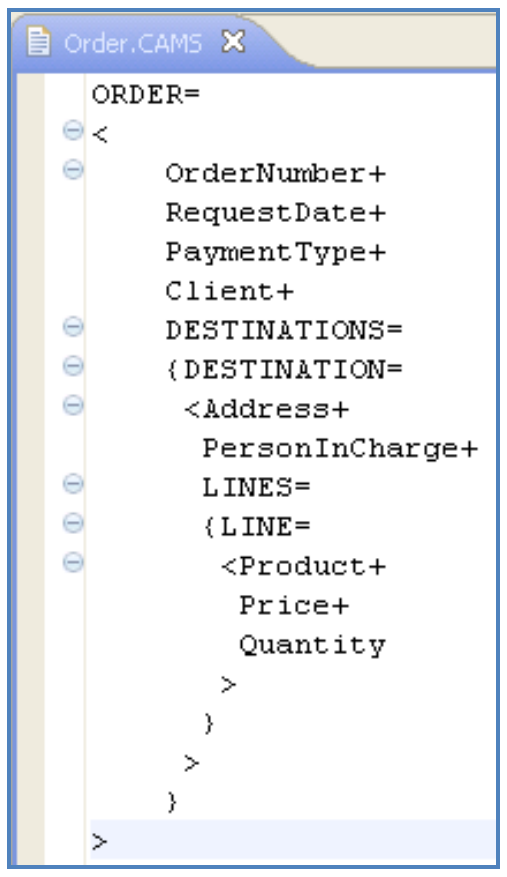

**b)** Ejemplo de Estructura de Mensaje

**Fig. 8.** Soporte para Estructuras de mensaje en el entorno Xtext

## 5.2. Soporte a las Estructuras de Mensaje mediante EMF

El paradigma DSDM puede dotar a los modelos de requisitos de un valor añadido: el potencial para derivar de ellos los modelos conceptuales que servirán para la generación automática de código. Con este fin, [Ruiz, España et al. 2010] define un proceso para integrar el Análisis de Comunicaciones en un entorno DSDM y presenta un metamodelo que especifica una parte del método. Para crear este metamodelo se hizo uso de las herramientas de modelado de UML de Eclipse Modeling Framework (EMF) y se generó el metamodelo Ecore correspondiente.

En este artículo se presenta una extensión del metamodelo de Análisis de Comunicaciones que permite el modelado de Estructuras de Mensaje. En la Fig. 9, las metaclases preexistentes aparecen con marco gris, mientras las metaclases añadidas al metamodelo se resaltan con marco negro.

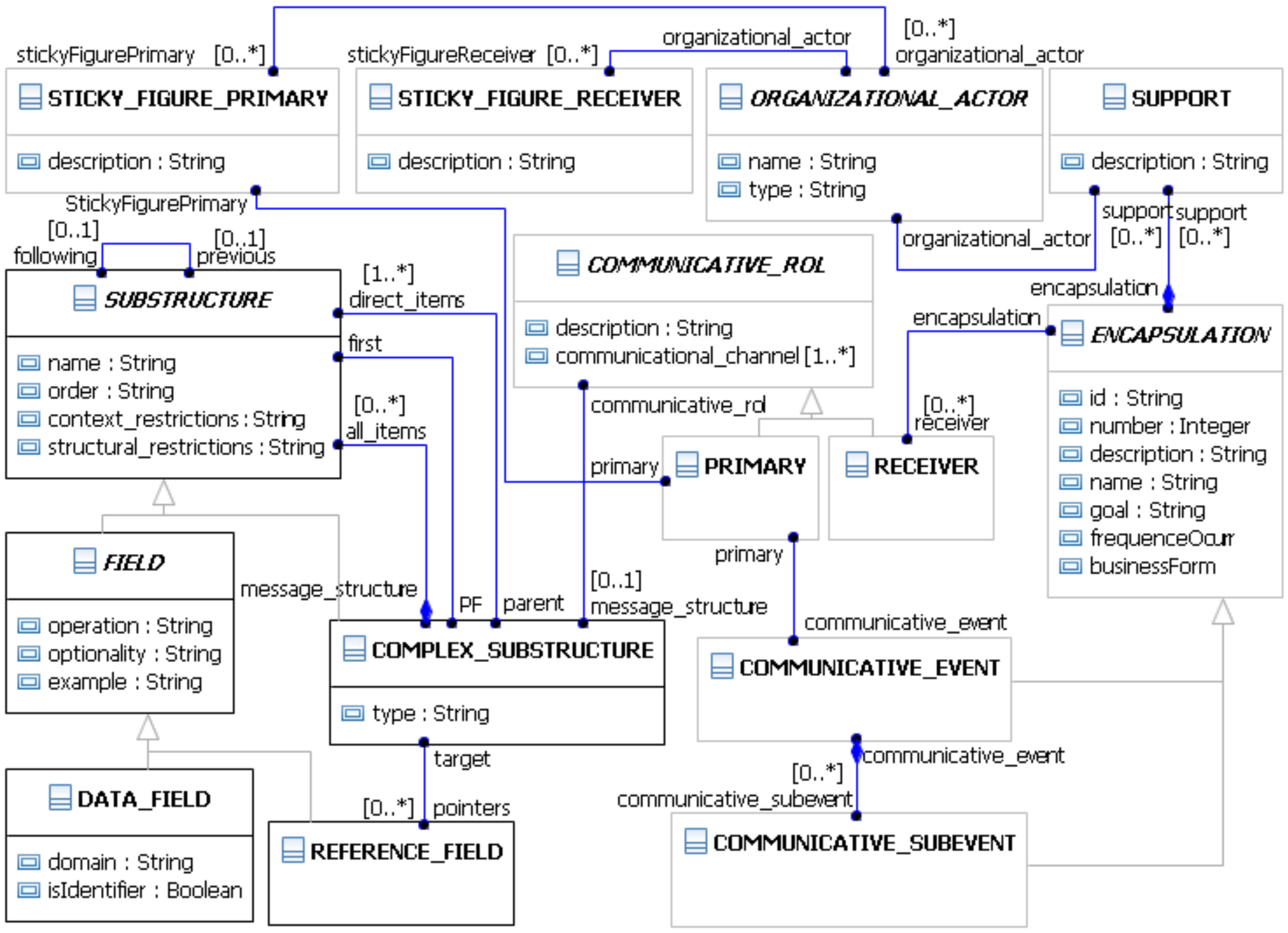

**Fig. 9.** Vista del metamodelo de Análisis de Comunicaciones extendido para dar soporte a las Estructuras de Mensaje

La Fig. 10 muestra la estructura de mensaje ORDER como una instanciación del metamodelo, en forma de árbol Ecore. El árbol representa gráficamente la composición de subestructuras complejas, quedando implícitos los operadores = y +. El tipo de cada subestructura se almacena en la propiedad Type de la metaclase COMPLEX_SUBSTRUCTURE (v.g. la solapa Properties muestra que la subestructura compleja DESTINATIONS es de tipo iteración).

Por un lado, la implementación en Xtext garantiza el cumplimiento de la gramática definida en EBNF para las Estructuras de Mensaje y ofrece un entorno de edición de estructuras de mensaje más eficiente y usable. Por otro lado, la implementación basada en EMF extiende la herramienta CASE para Análisis de Comunicaciones y su metamodelo Ecore ofrece la posibilidad de definir transformaciones modelo a modelo mediante lenguajes como ATL Transformation Language (ATL [Jouault and Kurtev 2005]) o Query/View/Transformation (QVT [OMG 2008]). En cualquier caso, ambas aproximaciones de implementación son complementarias, puesto que ambos entornos (Xtext y EMF) son integrables (esto se planea como trabajo futuro).

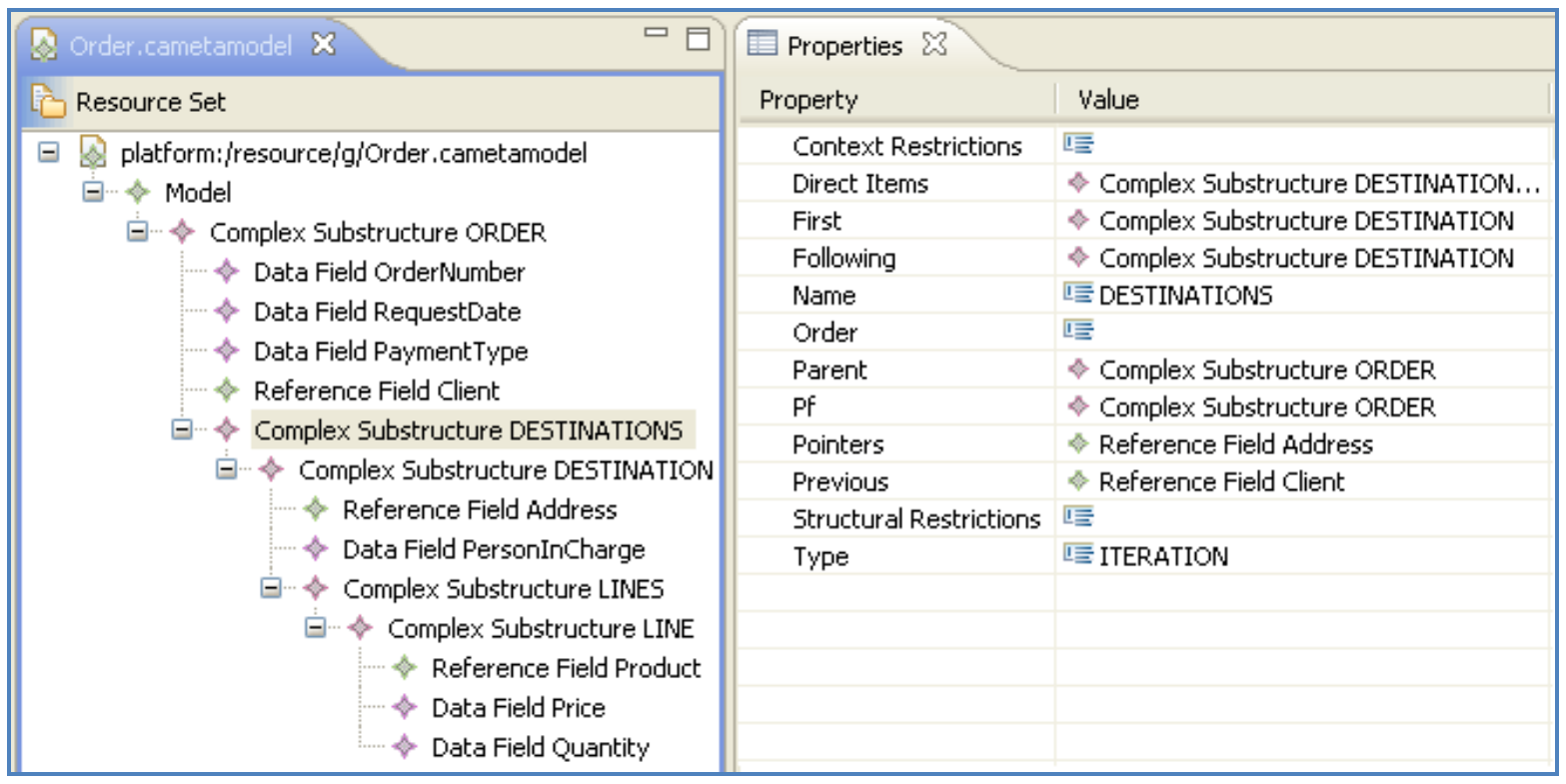

**Fig. 10.** Ejemplo de Estructura de Mensaje soportada en el entorno EMF

# 6. TRABAJOS RELACIONADOS

Existen varias aproximaciones comunicativas al análisis de sistemas de información. Por ejemplo, Action Workflow [Medina-Mora, Winograd et al. 1992], SAMPO [Auramäki, Lehtinen et al. 1988], Business Action Theory [Goldkuhl 1996], DEMO [Dietz 1999], Communication Analysis de Cronholm y Goldkunhl [Cronholm and Goldkuhl 2004], SANP [Chang and Woo 1994], Organisational Semiotics [Stamper 1997]. Compartimos con ellas la perspectiva comunicativa y numerosos fundamentos tomados de la teoría de la comunicación. Sin embargo, el Análisis de Comunicaciones difiere en varios fundamentos conceptuales, en la estructura de requisitos subyacente y en las guías para la modularidad de procesos de negocio. Véase [España, González et al. 2009] para una comparación más detallada. Adicionalmente, a diferencia de las aproximaciones mencionadas arriba, el Análisis de Comunicaciones enfatiza la especificación de los mensajes que corresponden a las interacciones comunicativas entrantes, para lo cual se proponen las Estructuras de Mensaje.

Un ancestro notable de las Estructuras de Mensaje es Backus-Naur Form (BNF). BNF es una notación para gramáticas libres de contexto propuesta por Backus y Naur en el desarrollo del compilador de Algol 60 [Backus, Bauer et al. 1963]. Los constructores gramaticales son la secuencia, representada mediante contigüidad, y la alternativa, representada mediante una barra vertical. Extensiones posteriores incorporan las llaves {,} para repeticiones y los paréntesis rectos [,] para la opcionalidad. El Análisis Estructurado de Sistemas adapta la notación BNF para la especificación de estructuras de datos [DeMarco 1979]. Fortuna et al. [Fortuna and Borges 2005] proponen extender Casos de Uso mediante con una notación basada en BNF para especificar interfaces informacionales, con una intención similar a las Estructuras de Mensaje.

Sin embargo, ninguna de las notaciones anteriores incluye un operador para parentizar explícitamente las secuencias. Esto impide describir subestructuras dentro de estructuras y obliga a fragmentar las estructuras complejas. Por ejemplo, la primera estructura (a) presenta una ambigüedad que solo se puede eliminar fragmentando la estructura en dos partes (b) o, como propone Estructuras de Mensaje, parentizando la agregación (c). Una ventaja de (c) sobre (b) es que (c) preserva la unidad del mensaje.

a) Vehículo=Matrícula+Marca+Modelo+Motor=Cilindrada+Válvulas+Combustible+Color
b) Vehículo=Matrícula+Marca+Modelo+Motor+Color
Motor=Cilindrada+Válvulas+Combustible
c) Vehículo=Matrícula+Marca+Modelo+Motor=<Cilindrada+Válvulas+Combustible>+Color

Otras técnicas que permiten modelar la interacción entre los actores y el SI pueden cumplir una función similar a las Estructuras de Mensaje; por ejemplo Diagramas de Secuencia UML o arboles de tareas como ConcurTaskTrees [Panach, España et al. 2008]. Sin embargo la experiencia nos ha demostrado la conveniencia de una notación más ligera, como las Estructuras de Mensaje, que tienen la ventaja de poder ser descritas textualmente.

# 7. DISCUSIÓN Y TRABAJOS FUTUROS

Los sistemas de información (SI) son soportes a la comunicación organizacional. En este artículo se presentan las Estructuras de Mensaje, una técnica que permite especificar las interacciones comunicativas entre los actores organizacionales y el SI. Además de detallar la gramática de las Estructuras de Mensaje, se explican sus usos en tiempo de análisis y de diseño del SI. En tiempo de análisis, facilitan la identificación de sucesos comunicativos (actividades del negocio que aportan nueva información significativa al SI) y complementan la descripción de los procesos de negocio. En tiempo de diseño, permiten derivar la memoria del SI y razonar el diseño de la interfaz. Se proveen guías metodológicas y ejemplos ilustrativos.

Las Estructuras de Mensaje forman parte del método de ingeniería de requisitos Análisis de Comunicaciones. Sin embargo, en el caso de que una organización desee continuar usando un método en particular, es posible extenderlo con Estructuras de Mensaje (esto ya se ha hecho para Casos de Uso); facilita su adopción el hecho de ser una técnica basada en otras ya conocidas (BNF, diccionarios de datos). Conviene hacer notar que, si las Estructuras de Mensaje se usan en combinación con otras técnicas para la especificación de la interacción (v.g. Diagramas de Secuencia), pueden aparecer redundancias. En este caso se aconseja la definición de reglas para derivar unos modelos a partir de otros, o establecer procedimientos para verificar la consistencia.

Un aspecto clave en el uso de las Estructuras de Mensaje es la distinción clara entre análisis y diseño. Este artículo provee de guías para el uso de la técnica en ambas etapas. Permiten prevenir que un analista especifique aspectos de diseño en tiempo de análisis, sin ser consciente de ello (algo que puede provocar la falta de experiencia con la técnica). Este riesgo de sobrecarga de trabajo y de sobre-especificación del modelo de requisitos disminuye con la práctica, al interiorizar la forma de uso de la técnica.

Al tratarse de una notación textual, las Estructuras de Mensaje ofrecen la ventaja de permitir la especificación mediante un procesador de textos o configurando campos textuales en herramientas CASE existentes. En cualquier caso, las experiencias en desarrollos industriales, docencia en másteres de ingeniería del software y experimentos de laboratorio (v.g. [España, Condori-Fernández et al. 2010]), nos han convencido de ofrecer un soporte más versátil. Este artículo presenta dos herramientas que soportan la técnica: una basada en la tecnología Xtext y otra basada en Eclipse Modeling Framework. Como trabajo futuro se planea integrar ambas herramientas para aprovechar la usabilidad de la primera y la capacidad para soportar transformaciones de modelos de la segunda. Actualmente se está trabajando en la derivación de modelos conceptuales a partir de modelos de requisitos que incluyen Estructuras de Mensaje, usando reglas de ATL Tranformation Language.